# Game Design Prototype with GIMs: Fostering Neurodiverse Connections through Storytelling

Yiqi Xiao[1[0009-0009-8725-9106]]

[1] University of Illinois Urbana-Champaign, Champaign, IL 61802, USA
yiqix3@illinois.edu

**Abstract.** This ongoing experimental project investigates the use of Generative Image Models (GIMs) in crafting a picture book creation game designed to nurture social connections among autistic children and their neurotypical peers within a neuro-affirming environment. Moving away from traditional methods that often seek to condition neurodivergent children to socialize in prescribed ways, this project strives to cultivate a space where children can engage with one another naturally and creatively through art and storytelling, free from the pressure to adhere to standard social norms. Beyond merely "story-choosing," the research highlights the potential of GIMs to facilitate "story-creating," fostering peer social connections in a creative and structured collaborative learning experience.

## 1 Research Background

### 1.1 The Social Needs of Autistic Children

Peer relationships, which include acceptance, friendships, and participation in peer networks as characterized by Gifford-Smith and Brownell [1], are crucial for children's well-being [2] and cognitive growth [3] and would influence their school performance [1]. Just like other children, research suggests that autistic children often have a strong desire for friendships [4]. However, multiple studies indicate that they frequently encounter social challenges [5, 6, 7]. Ghanouni et al. [8] note that many autistic children struggle with language, sensory processing, and recognizing social cues, which include but are not limited to picking up on jokes, maintaining eye contact during conversation, or appropriately interpreting others' facial expressions or tone of voice [9], and these disadvantages may lead to reduced or atypical interactions with peers. In other words, not "thinking and acting like others" in a traditional social setting prevents autistic children from forming friendships naturally with their neurotypical peers, which can have detrimental impacts on their overall development and well-being.

While the social interactions of autistic children might differ from those of neurotypical peers, this does not preclude their ability to build friendships. Recent research challenged the long-held beliefs that being autistic prevents them from forming friendships, revealing that many of them can form bonds with both neurotypical and autistic peers, albeit with differing interaction styles [10]. Recognizing the challenges that autistic children face in forming friendships within traditional social environments while acknowledging their readiness and capability to engage in social relationships, necessitates greater attention to their social needs and the obstacles they encounter.

### 1.2 Bonding between Autistic Children and Digital Games

Since autistic children usually face challenges in face-to-face communication, online spaces become an alternative for social expression and relaxation. Research shows that autistic children often show a pronounced inclination toward screen-based media, especially video games. Studies indicate that boys aged eight to eighteen on the autism spectrum average 2.4 hours per day gaming, based on a study of 169 individuals [11]. Barrington Campbell [12], an autistic adult, shared his thoughts about autistic people staying online: "If you are autistic, this can feel more than stressful (interacting with people face-to-face). Online gaming means autistic people can play in an environment we feel comfortable in and can control — if it’s not going well on screen, we can get into our safe spaces". Therefore, digital games present great opportunities to support the socialization of autistic children. By engaging with them through platforms they are comfortable with, designers and educators can create meaningful and effective interventions that align with autistic children’s preferences.

## 2 Related Work with AI

The development of AI technology has significantly transformed and advanced practices in the educational context, for example, AI can assist teachers in designing and delivering personalized, timely, iterative, and enriching educational experiences, with predictive insights, making education more effective and inclusive [13, 14, 15]. Compared to the broader field of education, the application of AI in supporting special education, especially for autism, is relatively new. However, there are still many insightful ongoing projects and proposals. The National AI Institute for Exceptional Education is developing AI screeners designed to detect speech and language processing challenges in young children by analyzing video and audio recordings of classroom interactions. They are also working on AI Orchestrators, a sophisticated multi-agent reinforcement learning framework that assesses each student's learning progression and recommends the most suitable interventions [16]. Similarly, a project uses the inbuilt accelerometer of a smartwatch to detect autism stereotypic behaviors, including hand flapping, painting, and sibbing, to better inform teachers and parents about their kids’ conditions [17]. In the field of serious games, the Film Detective game, designed by the AIVAS Lab at Vanderbilt University, helps autistic adolescents decode social situations and understand others’ emotions by interacting with and teaching a virtual agent, which makes learning enjoyable [18].

### 2.1 More to Explore in Design with AI

While these projects employ various approaches to assist autistic individuals, they share some similarities. Many current approaches consider AI as an extension of the role of educators [18], caregivers [29], and other human experts [16, 17] aiming to achieve previously unimaginable efficiency. Although all these practices are valuable, they typically involve AI replicating existing adult roles without changing or enriching the forms of assistance available to children. An approach worth further exploring is using AI to empower children following their own wishes and creativity. While their capabilities are increased, their agency is also bolstered. This opens up more possibilities for children's learning and growth.

Another observation is that most research focuses on text, image, video analysis, and text generation, rather than image generation [16, 17, 18, 29]. This may be related to the previous fact that people typically use AI to replace adults' work with children while drawing pictures might not be the main part. It is important to address that, Generative Image Models (GIMs), a branch of generative AI technology designed to create images from natural language descriptions, have significant potential to contribute to special education and serious game design.

### 2.2 Opportunities that Generative Image Models Bring

In the field of autism education, image generation could offer tremendous opportunities for autistic children, especially in communication and social interaction. Many autistic children are visual thinkers. Dr. Temple Grandin [19, 20, 21], an autistic adult and advocate, wrote that many autistic individuals, including herself, think in pictures very efficiently [19]. Research also shows that some autistic children demonstrate high levels of visual-spatial skills [20] and visual memory [21]. These alternative strengths are sometimes revealed by young children with autism during the creative art-making process [22]. In these cases, GIMs can be an excellent tool for fostering visual expression and communication between autistic children with their peers, without being constrained by language and drawing skills.

In the context of "fair use" in education, AI-generated artwork can significantly accelerate game production and reduce labor costs. This enables educators and game designers with a forward-thinking vision to create without being limited by technical constraints. More importantly, AI could introduce new possibilities for game mechanics. Games with storytelling elements like Heavy Rain [23] and Detroit: Become Human [24] have fascinating storylines and visuals but still could only allow players to "choose" their path within pre-designed routes. With AI, players are allowed to "create" their own adventures. The capability for them to "create" the characters they are going to encounter, the tools that could help them solve problems, and the subsequent events in each playthrough will greatly increase the playability of the games.

# 3 Project Introduction

## 3.1 Research Question

How might we create an experience that fosters bonding among neurotypical and neurodivergent school-age children with the assistance of Generative AI?

## 3.2 Project Overview

This project is developing a hybrid learning experience designed to engage both autistic children and their neurotypical peers in a two-phase interactive process - creating a digital picture book online and telling the story together in person. The story setting is inspired by the well-known "Jack and Jill" rhyme, a narrative that depicts two children embarking on an adventure and experiencing the world together without relying on verbal communication. This story resonates with the concept of equal and diverse connections—a crucial idea to be emphasized when supporting autistic children in social scenarios. In the game, players first create and name their avatars, and start their adventure of fetching water by making a series of choices. The intention is to relate simple visual choices with realistic consequences, encouraging children to engage deeply with the narrative they're crafting. Some of the decisions will further impact the narrative's outcome, such as the location of the water source. All their decisions will be recorded and result in a digital storybook with a cover that showcases their avatars and includes their names as the title.

Upon completing the picture book activity, the digital book will be printed and sent to each child's home or classroom if they are classmates. This not only offers the children a tangible keepsake of their joint effort but also serves as a tool or ‘seed’ for further creative interaction. They can tell and even perform the story together under the guidance of the instructor, enhancing their collaborative experience.

## 3.3 Design Principles

This story-creating environment is built on three fundamental design principles.

- Firstly, it forgoes text and judgment. This work aims to create a space free from social pressure or fear of failure, enabling autistic children to express themselves and connect with others confidently.
- Secondly, it offers a visual experience that extends beyond mere images. With art elements generated by GIMs representing various objects, weather, actions, intentions, and sensations, players are invited to explore the implications of each visual choice and consider how it influences the story, for example, they choose the sky color to indicate the time of day, select characters to encounter (Fig. 1), and employ visual symbols in the dialogue box that influence the storyline.
- Thirdly, it is all about creation. Instead of "choosing" their adventure, players are "creating" their adventure. While initially, players may seem to be able to pick from pre-generated options as the game develops, it's anticipated to evolve, allowing AI to assist children directly, which will expand the possibilities for children to freely create their journeys, transforming the game into a genuinely open-ended experience.

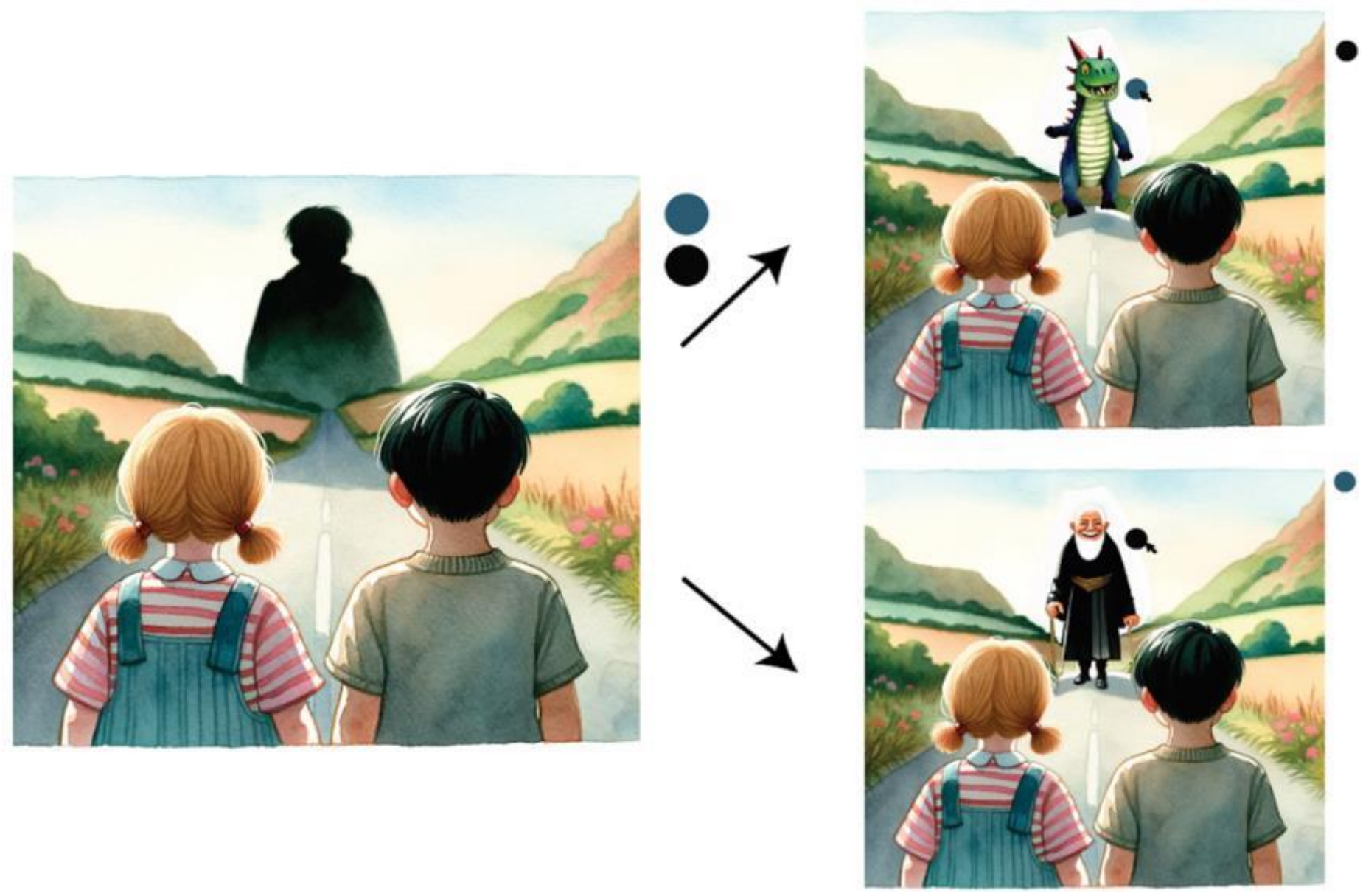

**Fig. 1.** Players choose which character to encounter

### 3.4 Making All the Art Elements Using Generative AI

Utilizing generative AI for artwork creation has significantly enriched the game development process and has brought satisfying results. DALL·E [25], Midjourney [26], and Adobe Photoshop (Beta) [27] have been instrumental in producing the initial image prototypes and will further support players in generating unique story elements each time they play. Additionally, the author has engaged with ChatGPT [28] for creative insights on narrative development and intends to further leverage its capabilities to assist with the game's coding phase.

### 3.5 Data Collection

Qualitative data will be collected from three sources: on-site observations of players' behavior while interacting with the game, the digital picture books completed by users online, and the observations or recordings of their telling the story they made together (Fig. 2), Additionally, follow-up questionnaires will be given to the caregivers of the participating children. These questionnaires will inquire about the children's experiences post-game, including questions like how many times they shared their picture book with others, how often they mentioned this experience, and whether they expressed a desire to design another book using this game. This approach aims to understand the children's experiences comprehensively, facilitating a thorough evaluation of the project. This evaluation is crucial for the project's future refinement.

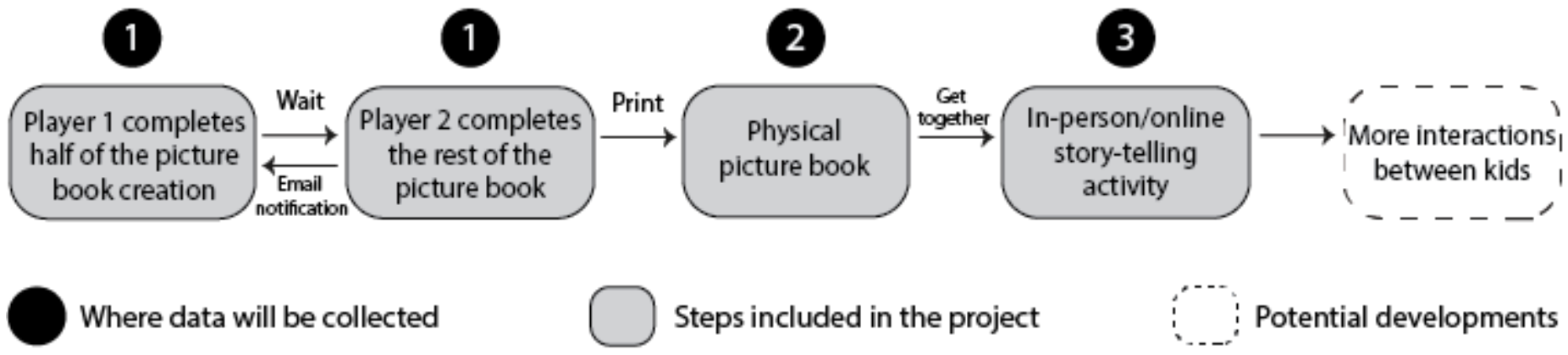

**Fig. 2.** The process of the project

## 4 Discussion

While it is designed for autistic children with social challenges, it can also benefit other neurodivergent children who encounter the same issue, as well as neurotypical children who are more inclined towards artistic creation compared to engaging in conventional social settings with their peers. Besides, the process collects extensive data about children's decision-making during gaming, which can potentially help scholars in neuroscience, education, and psychology to gain insight into the children's behaviors such as self-expression, creativity, learning, and social interactions.

Currently, this project is in the prototype stage, with all art elements pre-generated during the game design and then offered as choices to players. This game will keep evolving and explore the possibility of enabling children to interact directly with AI to generate their desired stories. This approach requires not only technical optimization but also careful consideration of safety. It is hoped that this work will spark further discussions about the application and safety of Generative Image Models (GIMs), aiding in its' development.

## 5 Conclusion

This paper identifies the potential of GIMs in game design for autism education and proposes a prototype of a digital picture book-making game using GIMs. This game assists in creating a hybrid, structured, and collaborative learning experience between autistic children and their non-autistic peers. The digital platform of this game serves as a space that fosters their cocreation. A picture book they both contribute to allows autistic children to comfortably enter social interactions with peers. This innovative game-based approach fosters neurodiverse communication by providing a structured yet flexible environment where autistic children can creatively build peer relationships through creating and telling a story together.

**Acknowledgments.** The author would like to thank Juan Salamanca Garcia and Peizhen Wu for their valuable assistance in refining the language and improving the clarity of this paper.